\documentclass[twocolumn,pra,superscriptaddress,10pt,a4paper,showpacs]{revtex4}
\usepackage{graphicx,amsmath,amsfonts,algorithm}
\usepackage{graphics}
\usepackage{amssymb}

\usepackage{color}
\usepackage{bbm}

\def\ket#1{| #1 \rangle}

\newtheorem{prop}{Proposition}\def\PRO{\begin{prop}}\def\ORP{\end{prop}}
\newtheorem{coro}{Corollary}\def\COR{\begin{coro}}\def\ROC{\end{coro}}
\newtheorem{theo}{Theorem}\def\TH{\begin{theo}}\def\HT{\end{theo}}
\def\TH{\begin{theo}}\def\HT{\end{theo}}
\newtheorem{defi}[prop]{Definition}\def\DE{\begin{defi}}\def\ED{\end{defi}}
\newtheorem{lemme}[prop]{Lemma}\def\LE{\begin{lemme}}\def\EL{\end{lemme}}

\def\EQ#1{\begin{eqnarray}#1\end{eqnarray}}

\def\qed{$\Box$}
\newcommand{\djj}{d\kern-0.4em\char"16\kern-0.1em}

\begin{document}
 \title{Truly noiseless probabilistic amplification}
\author{Vedran Dunjko} 
\affiliation{SUPA, School of Engineering and Physical Sciences, Heriot-Watt University, Edinburgh, UK}
\affiliation{Division of Molecular Biology, Ru\djj er Bo\v{s}kovi\'{c} Institute, Bijeni\v{c}ka cesta 54, P.P. 180, 10002 Zagreb, Croatia}
\author{Erika Andersson}
\affiliation{SUPA, School of Engineering and Physical Sciences, Heriot-Watt University, Edinburgh, UK}

\begin{abstract}
Most of the schemes for ``noiseless" amplification of coherent states, which have recently been attracting theoretical and experimental interest, share a common trait: the amplification is not truly noiseless, or perfect, for non-zero success probability. While this must hold true for all phase-independent amplification schemes, in this work we point out that truly noiseless amplification is indeed possible, provided that the states which we wish to amplify come from a finite set. Perfect amplification with unlimited average gain is then possible with finite success probability, for example using techniques for unambiguously distinguishing between quantum states. Such realizations require only linear optics, no single-photon sources, nor any photon counting. We also investigate the optimal success probability of perfect amplification of a \textit{symmetric} set of coherent states. There are two regimes: low-amplitude amplification, where the target amplitude is below one, and general amplification. For the low-amplitude regime, analytic results for the optimal amplification success probabilities can be obtained. In this case a natural bound imposed by the ratio of success probabilities of optimal unambiguous discrimination of the source and amplified states can always be reached. We also show that for general amplification this bound cannot always be satisfied. 
\end{abstract}
\pacs{03.67.-a 
42.50.Dv 
42.50.Ex 
}

\maketitle

\section{Introduction}
There has recently been widespread theoretical and experimental interest in schemes for ``noiseless" amplification of coherent states~\cite{Grangier, Xiang, Usuga, Zavatta, Croke, Jeffersampl}. These schemes aim to implement the operation $|\alpha\rangle \rightarrow |g\alpha\rangle$, for $g>1$ and any $\alpha$. This is not possible to achieve perfectly with unit probability, but can be done probabilistically with arbitrarily high fidelity. Noiseless amplification could for example be used in quantum repeaters, or for entanglement purification through ``breeding" larger Schr\"odinger cat states from ``kittens", by probabilistically transforming $N_\pm(|\alpha\rangle\pm|-\alpha\rangle)$ into $N'_\pm(|g\alpha\rangle\pm|-g\alpha\rangle)$ with high fidelity.

Common to all existing schemes is that the amplification is not truly noiseless, or perfect, for non-zero success probability. That is, the fidelity approaches unity only in the limit of vanishing success probability.
This must in fact hold for any phase-independent amplification scheme~\cite{Croke}. The suggested schemes achieve higher fidelity for smaller $\alpha$ or smaller gain, but it is only if either $|\alpha\rangle=|0\rangle$ or $g=1$ that the fidelity can be 100\% for non-zero success probability, in which case of course no amplification actually takes place. For experimental realisations, the overall success probability is usually not even quoted, and only the fidelity in case of successful operation is reported as a figure of merit. A complete and fair comparison of the different schemes is therefore difficult.
The success probability, especially for schemes that involve single-photon states as resources,
is nevertheless usually very low.

In this paper we want to point out that in contrast to existing theoretical and experimental schemes, there {\em is} in fact a way to achieve truly noiseless amplification, that is, 100\% fidelity,
also for finite non-zero success probability and finite non-zero coherent state amplitudes. This is possible if one relaxes the demand that the amplification should work 
for any $|\alpha\rangle$, and instead selects any finite number of coherent states that one wants to amplify perfectly. The restriction to a finite set of states need not be serious, since many 
quantum information and communication protocols use a selected set of states, including quantum cryptography~\cite{BB84,barbosa-2002,Leuchs},
 blind quantum computing~\cite{10.1109/FOCS.2009.36}, and quantum digital signatures using coherent states~ \cite{PhysRevA.74.022304, ClarkeDigSign}. 
For example, the set of symmetric coherent states $|\alpha e^{im 2\pi/N}\rangle$, where $\alpha$ is fixed and $m=1,2, \ldots N$, may be amplified truly perfectly with non-zero success probability. 

In fact, any set of linearly independent quantum states, coherent or other, may be amplified or cloned perfectly with a finite non-zero probability of success. This follows from the fact that linearly independent states may be unambiguously distinguished from each other with finite success probability~\cite{Chefleslin}. Perfectly identifying a quantum state clearly allows us to fabricate an unlimited number of copies, or equivalently, to prepare a state with the same phase and arbitrarily high amplitude. Hence, it is not only possible to perfectly amplify any linearly independent set of states, but the average gain of truly noiseless probabilistic amplification can be {\em arbitrarily high}, since the success probability times the gain 
is unlimited. 
Moreover, unambiguous state discrimination of coherent states may be realized using only linear optics and non-photon-number-resolving photodetectors, without using auxiliary non-classical states~\cite{Huttnertwocoh, vanEnk}. The same resources allow also realization of perfect amplification based on unambiguous state discrimination (USD).

When discussing amplification, the so-called classical linear amplifier is often used as a benchmark \cite{Zavatta}. 
This is a measure-and-prepare approach to amplification or cloning, where the state is first estimated and based on this the amplified state prepared. Depending on what states we wish to amplify or clone, however, the optimal measure-and-prepare classical amplifier protocol will be different. Existing amplification protocols for coherent states are phase-independent \cite{Grangier, Xiang, Usuga, Zavatta, Croke, Jeffersampl}, or consider some other continuous distribution of coherent states \cite{Cochrane, Sabuncu,Muller,Sacchi,Guo2}.
For a continuous input distribution, the realized amplification fidelity can never be perfect.
In contrast to this, we consider a restricted setting where the inbound set of states is finite and linearly independent. This will be related to a {\it different} measure-and-prepare protocol than phase-independent amplification, in other words, to a different classical amplifier. 

If we do not require arbitrarily high gain, then the success probability can be higher than for schemes based on unambiguous state discrimination. For amplification of symmetric sets of coherent states, results on transforms between sets of symmetric states~\cite{st} are key to working out what processes are possible.
Such transforms might be termed ``umbrella transforms", if we visualize the symmetric states as the spines of an umbrella in a space of suitable dimensionality. A probabilistic transform that decreases pairwise overlaps, one example being noiseless amplification, may then be thought of as ``opening the umbrella".
We will be concerned with the theoretical limits of limited-gain perfect amplification of a restricted set of possible input states, in particular, the optimal success probabilities of such transforms.

The paper is organized as follows. In Section~\ref{sec:linopt}, we briefly review unambiguous discrimination of coherent states using linear optics, and discuss how to use this for truly noiseless amplification. Definitions related to transformations between sets of quantum states are given in Section~\ref{sec:pre}. In Section~\ref{sec:ampl}, we investigate truly noiseless amplification of coherent states, for finite gain, by viewing it as a transform between symmetric sets of states. As already mentioned, the success probability can then be higher than for procedures that use state discrimiation. It turns out that there are two regimes; small amplitude amplification, where the amplitudes of both initial and amplified states are below one, and a general regime where the amplitude of the final states, or of both initial and final states, are above one. 
As shown in~\cite{st}, transforms between sets of states may be ``leaky" or ``leakless", depending on whether there is an extra ``leak" state correlated with the desired output in the case of success. It turns out that in the small amplitude regime, the optimal ``umbrella transforms" for noiseless amplification are leakless, whereas in the general regime, they may be leaky.
We finish with a discussion.

\section{Amplification of coherent states using linear optics}
\label{sec:linopt}

Ivanovic~\cite{Ivanovic}, Dieks~\cite{Dieks} and Peres~\cite{Peres} realized that  two non-orthogonal quantum states can be unambiguously distinguished from each other with a certain probability. That is, if the measurement succeeds, the result is always correct, but there is a chance that the measurement fails, giving an inconclusive result. The failure probability for the optimal procedure is equal to the overlap between the two quantum states. In the completely general case, optimal unambiguous measurements are not easy to find analytically~\cite{SteveSarahReview, JanosReview},  but such a measurement is at least possible as soon as at least one of the quantum states is linearly independent of the others~\cite{Chefleslin}.

For two coherent states $|\alpha\rangle$ and $|-\alpha\rangle$, the optimal measurement may be realized using only linear optics~\cite{Huttnertwocoh}. The state to be identified, $|\pm\alpha\rangle$, is directed onto a balanced beam splitter, with a fixed state $|\alpha\rangle$ incident on the other input port. If the phase relationships between output and input ports are arranged so that the beam splitter transforms $|\alpha\rangle_1\otimes|\beta\rangle_2$ to $|(\alpha+\beta)/\sqrt{2}\rangle_1\otimes|(\alpha-\beta)/\sqrt{2}\rangle_2$,
we see that if the state to be identified was $|\alpha\rangle$, then output port 1 will contain $|\sqrt{2}\alpha\rangle$ and port 2 will be empty, and if it was $|-\alpha\rangle$, then output port 1 will be empty and output mode 2 contain $|-\sqrt{2}\alpha\rangle$.   By detecting photons in the output ports, we can therefore unambiguously tell whether the state in input port 1 was $|\alpha\rangle$ or $|-\alpha\rangle$. 
Since any coherent state contains a vacuum component, we may not see any photons at all, which corresponds to the inconclusive outcome. The probability for this is $\langle 0|\sqrt{2}\alpha\rangle = \langle -\alpha|\alpha\rangle = \exp(-|\alpha|^2)$, which is the optimal (minimal) failure probability. Clearly, no photon counting is required, only being able to tell the difference between the vacuum and any nonzero number of photons.

For a balanced beam splitter with other phase relationships, we can adjust the phase of the fixed state in input port 2 so that the procedure still works. Also, if the two states to be distinguished are not $|\pm\alpha\rangle$ but $|\alpha\rangle$ and $|\beta\rangle$, then we can precede the described measurement with displacement of the unknown input mode, containing either the state $\vert \alpha \rangle$ or $\vert \beta \rangle$, by $-(\alpha+\beta)/2$  using a beam splitter, and then distinguish $|\pm(\alpha-\beta)/2\rangle$ using the technique above. 

\begin{figure}
\includegraphics[scale=0.45,viewport=0 400 700 700,clip]{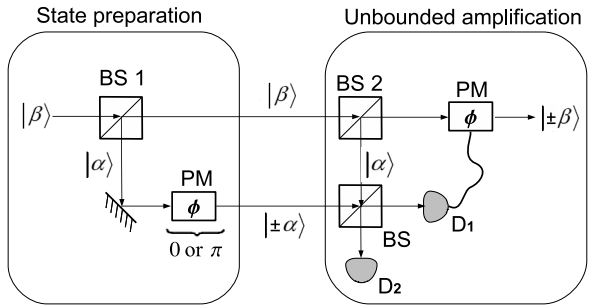}

\caption{Truly perfect amplification of the states $|\pm\alpha\rangle$ based on unambiguous discrimination, where we assume that $|\beta|\gg|\alpha|$. 
The beamsplitters denoted $\mathsf{BS\ 1}$ and $\mathsf{BS\ 2}$ split off a minor fraction of the strong beam $\ket{\beta}$ of amplitude of norm $|\alpha|$.
The beamsplitter $\mathsf{BS}$ is balanced, and boxes labeled $\mathsf{PM}$ denote phase modulators. 
The amplification procedure fails only if both detectors $\mathsf{D}_1$ and $\mathsf{D}_2$ fail to detect a photon.}
\label{fig:twoampl}
\end{figure}

This unambiguous measurement may be used for perfect amplification as shown in Figure~\ref{fig:twoampl}, where the first box shows a suggested way to prepare the states to be distinguished, and the second box the unambiguous measurement itself . The fact that we need to specify the phases of $|\pm\alpha\rangle$ implies that there exists a phase reference beam, which we without loss of generality assume to be $|\beta\rangle$, where $\alpha$ and $\beta$ have the same phase, but different amplitude; a strong reference beam would have $|\beta|\gg |\alpha|$. The fixed state $|\alpha\rangle$ in input mode 2 is likely also split off this reference beam, as shown in Fig.~\ref{fig:twoampl}. Conditional on whether the state is identified as $|\alpha\rangle$ or $|-\alpha\rangle$, we implement the corresponding phase shift on the reference beam, giving the amplified state. The gain is then only limited by how strong the reference beam is.
Alternatively, we could amplify relative to some other reference beam, not necessarily with the same phase as $|\alpha\rangle$ (but we still need the fixed state $|\alpha\rangle$ with the correct phase in input mode 2 for the unambiguous measurement).

A similar procedure is possible for distinguishing between more than two coherent states using linear optics, but will then not attain the optimal success probability~\cite{vanEnk}. In short, if there are $N$ possible different states, then we can split the unknown state in $N$ beams using a multiport, and test each component against one of the possible states (with amplitude suitably scaled down) using a beam splitter similar to as described above for two coherent states. If we manage to rule out all but one of the possible states, then we have unambiguously identified the input state as the remaining one.
(Actually, we would only need to split the state to be identified in $N-1$ components, since if we manage to rule out all but one of the possible states, then the state must have been the remaining one.) The success probability of this procedure is non-zero, but not optimal. It can be somewhat improved by splitting the original state in $M$ copies, with $M\rightarrow\infty$, still using only linear optics~\cite{vanEnk}.

Any such procedure to unambiguously distinguish a finite number of coherent states may be used to noiselessly amplify them with a finite success probability and gain only limited by the strength of a reference beam, similar to amplification of two coherent states illustrated in Fig.~\ref{fig:twoampl}. If we manage to identify the state, we implement the corresponding phase shift on the reference beam.
Although this requires only linear optics and detectors that do not resolve photon numbers, the disadvantage is that such a procedure cannot be used to amplify a superposition of the possible incident states. This obviously limits the usefulness when the superposition is important, such as when ``breeding" larger cat states in order to enhance entanglement.

It is nevertheless in principle possible to realize truly noiseless amplification in such a way that superpositions are preserved. This is because one can in principle realize
unambiguous state discrimination 
in two steps. First, one probabilistically transforms the selected set of non-orthogonal states into orthogonal ones without destroying possible superpositions of the states in the set, and this is followed by a measurement to distinguish the different orthogonal states. Truly noiseless amplification that preserves superpositions can then be achieved by omitting the final measurement, and only registering whether the first step succeeded or failed (how this works is also clarified by Eq. \ref{critU} in the following Section). 

If the states to be amplified are symmetric to start with, then it follows from results in~\cite{st} that the success probability can be made independent of the initial state, and therefore the weights of the states in the superposition will be preserved. If the set of states is not symmetric, then the success probabilities for different states in the ``source" set may not be equal. Amplification that preserves the superposition but re-weights the individual states is then still possible. Also, if we base the procedure on unambiguous state discrimination, then the amplified states in the superposition will be orthogonal, corresponding to infinite coherent state amplitude (the unambiguous measurement, if completed, would give us perfect knowledge about which state was prepared, if only one of the initial states 
was prepared). Alternatively, if the amplitude of the amplified states is below one, then amplification that preserves superpositions is also possible, since then a leakless transform is possible, as we show in Sec.~\ref{sec:smallamp} (the concept of leak is introduced in the next section).
We leave it open whether superposition-preserving truly noiseless amplification of coherent states could be realized using only linear optics. 

Alternatively, we could remove the detectors and the strong reference beam $|\beta\rangle$ in the second box in Fig.~\ref{fig:twoampl}, and view the state exiting from the beam splitter labelled ``\textsf{BS}", after combining the incident state $|\pm \alpha\rangle$ with the fixed state $|\alpha\rangle$, as an ``amplification", with gain $\sqrt{2}$, of the incident state. This ``amplification" occurs with unit probability, that is, it is deterministic, and transforms a superposition $N_\pm(|\alpha\rangle_1 \pm |-\alpha\rangle_1)$ into $N'_\pm(|\sqrt{2}\alpha\rangle_1\otimes|0\rangle_2 \pm |0\rangle_1\otimes|-\sqrt{2}\alpha\rangle_2)$.
The deterministically ``amplified" state will then be a superposition of different output modes. 
However, the overlap between the incident states $|\pm\alpha\rangle_1\otimes |\alpha\rangle$ is necessarily the same as the overlap between the states $|\sqrt{2}\alpha\rangle_1\otimes|0\rangle_2$ and $ |0\rangle_1\otimes|-\sqrt{2}\alpha\rangle_2$. Thus it is questionable if this process really could be called amplification, without subsequently combining the amplified states into the same spatial mode. That can only be done probabilistically, since otherwise we would be able to deterministically decrease the overlap of two quantum states, which is impossible.

Amplification with in principle unlimited gain will necessarily have the same optimal success probability as unambiguous state discrimination. 
We will now proceed to investigate the optimal success probabilities and other properties of the amplification transforms for specified finite levels of gain. For this, we first need to state some definitions related to transforms between sets of quantum states.

\section{Transforms between sets of states}
\label{sec:pre}
In the previous section, we considered perfect amplification with unlimited gain, based on USD techniques. We noted that if USD is viewed as a heralded transformation from non-orthogonal to orthogonal states, followed by a measurement to distinguish between these states, there is nothing forbidding amplification which also preserves superpositions. In contrast to this, we will now consider perfect amplification of a finite set of coherent states, with {\it limited} gain. Our approach will be based on viewing amplification as a transformation between finite sets of states, and our goal to find the limits of success probability for such transforms, \emph{i.e.}, for truly perfect amplification.

We consider two sets of $N$ pure states, called the \textit{source} and \textit{target} sets, denoted (respectively)
\begin{eqnarray}
A=\{|a_i\rangle \} ; \ B=\{|b_i\rangle\}, \nonumber
\end{eqnarray}
and a heralded probabilistic transform $\mathcal{T}$ 
which for input state $|a_i\rangle$ produces the state $|b_i\rangle$ with probability $p_i$, and a $|Fail\rangle$ state with probability $1-p_i$.
By Theorem 3 in \cite{chefles-2003} such a  transform exists iff there exists a unitary transform $U$ performing
\begin{equation}
\label{critU}
U |a_i\rangle = \sqrt{p_i} |b_i\rangle\ket{\psi_i}|0\rangle + \sqrt{1-p_i} |Fail\rangle|\phi_i\rangle|1\rangle,\ \forall i
\end{equation}
for some sets of states $L=\{\ket{\psi_i}\}_N$ and $R=\{|\phi_i\rangle \}_N$. The states $|0\rangle$ and $|1\rangle$ are orthogonal. To complete the realization of $\mathcal T$, after the application of $U$ the third register is measured in this basis, and, optionally, the second register may be traced out. 

When the transform succeeds, the state $|b_i\rangle$ is generated along with a state $\ket{\psi_i}$, possibly correlated with the input state. This state leaks additional information about $i$, hence the set $L$ is called the \textit{leak}. 
When the transform fails, the constant state $|Fail\rangle$ is produced along with the state $|\phi_i\rangle$ which may be correlated with the source state, and may be used to attempt a reconstruction of the target state $|b_i\rangle.$ This set of states $R$ we call the \textit{redundancy}.
The leak (redundancy) states are not correlated with the input state iff  the states in the leak (redundancy) are identical for all source states, up to global phase.
If the success probabilities do not depend on the input state, the transform is called \textit{uniform}.
For uniform transforms (of success probability $p$) the criterion (\ref{critU}) may be rewritten, in terms of the Gram matrices of the sets $A$,$B$, $L$ and $R$ respectively, as
\begin{equation}
\label{critG}
G_A = p\ G_B \circ G_L + (1-p)\ G_R,
\end{equation}
where $\circ$ denotes the Hadamard (point-wise) matrix product.  The Gram matrix of a set of states $\{|c_i\rangle \}$ is defined as the square matrix with elements $\langle c_i| c_j\rangle$.

Finally, a finite set of states is \textit{symmetric} if there exists a fixed unitary which, when applied on the $i^{th}$ state, produces the $(i+1 \mod N)^{th}$ state.
Symmetric states are interesting as they often appear in quantum protocols (e.g. many quantum key distribution schemes \cite{BB84,barbosa-2002,Leuchs} and in blind quantum computing~\cite{10.1109/FOCS.2009.36} and quantum digital signature schemes with coherent states~\cite{PhysRevA.74.022304, ClarkeDigSign}).

\section{Amplification as state transforms}
\label{sec:ampl}

If the source set of coherent states we wish to amplify perfectly is known, and the required gain $g>1$ is pre-set, then the amplification procedure becomes a particular type of state transform which has been studied in \cite{st}. Here, we will assume that the source set of coherent states is a symmetric  set of $N$ states. The source and target states are then
\EQ{
\label{source}
A = \left\{\ket{a_i}:= \ket{e^{i \theta_k} \alpha  }\right\}_{k=0}^{N-1}, 
B = \left\{\ket{b_i}:= \ket{ e^{i \theta_k} \beta   }\right\}_{k=0}^{N-1},}
where $\theta_k = {2 k \pi i}/{N}$ and $\beta = g\alpha.$
An amplification transform takes states from set $A$  to corresponding (amplified) states in set $B$, and without the loss of generality we define the amplitudes $\alpha$ and $\beta$ to be real positive numbers.
The question is with what success probability such amplification is possible.

Since the set $A$ is a set of linearly independent states, using state discrimination one can always perform a measure-and-prepare procedure, and in fact reach any desired, unlimited gain. Thus, the lower bound on the success of an amplification procedure is given by $d_A$, denoting the success probability of unambiguous state discrimination of the states in $A$.
If we also take into the account the probability of unambiguous discrimination of states in $B$, an upper bound of the success probability of amplification can be derived. If $d_A$ and $d_B$ are the respective probabilities of optimal unambiguous discrimination of states in the sets $A$ and $B$,
then the corresponding amplification transform cannot succeed with a probability higher than
\EQ{
p_{up}=\frac{d_A}{d_B}\label{up}
} as a higher success probability would violate the optimality of $d_A$.
Similar methods have been used to bound the success probability of decreasing the overlap of two quantum states, which includes state-dependent cloning or two states~\cite{chefles-1998-31}. 
Similarly, one could derive other bounds by observing the optimal probabilities of minimum error measurements~\cite{Helstrom} on the sets $A$ and $B$, or in fact any measurement optimizing any other figure of merit (e.g. maximal mutual information, maximum likelihood, etc.). 

As we will show, the bound $p_{up}$ can in fact be reached for source and target amplitudes below one, whereas for target state amplitudes above one it cannot always be saturated.
The techniques we use have been developed in \cite{PhysRevA.65.052314,chefles-2003, st}.
By the results given in \cite{PhysRevA.65.052314,chefles-2003}, an amplification transform succeeding with probability $p$ exists if the equality given in equation (\ref{critG}) is satisfied for some Gram matrices of states \footnote{A matrix is a Gram matrix of states if and only if it has unity across the diagonal and is positive semi-definite, see \cite{chefles-2003}.} $G_L$ and $G_R$, and where $G_A$ and $G_B$ are the Gram matrices of the source and amplified coherent states, respectively.
Since $A$ and $B$ are symmetric sets of states, the matrices $G_A$ and $G_B$ are circulant \footnote{A circulant matrix is a square matrix, whose $i^{th}$ row is the right cyclic shift of the $(i-1)^{st}$ row.}, and hence diagonalize in the unitary discrete Fourier transform basis which is given by the columns or rows of the unitary discrete Fourier matrix of appropriate size $N$, 
\begin{eqnarray}
uDFT=1/\sqrt{N} \left[ \exp \ \dfrac{-2 \left(p-1\right) \left(q-1 \right) i \pi}{N }         \right]_{p,q}.
\end{eqnarray} 
 Moreover, by Lemma 4 in \cite{st},  if there exists any amplification procedure for symmetric states succeeding with some success probability, then there exists an amplification procedure succeeding with the same success probability, where the leak and redundancy are symmetric sets of states.

Thus, in order to find the optimal success probability, we may assume that all the matrices appearing in the existence criterion (\ref{critG}) are circulant, and they all diagonalize in the unitary discrete Fourier transform basis. Criterion (\ref{critG}) may then be written in terms of vectors containing the eigenvalues of the Gram matrices as
\EQ{
\lambda_A = p \lambda_B \ast \lambda_L + (1-p) \lambda_R. \label{critEvs}
} 
In this expression the vector $\lambda_X$ contains the diagonal elements of the matrix $uDFT^\dagger.G_X.uDFT$ which is diagonal when $X$ is a symmetric set of states, and $\ast$ denotes the circular convolution of vectors, defined component-wise as 
\EQ{
(\lambda_B \ast \lambda_L)_i = \frac{1}{N}\sum_{j=0}^{N-1} (\lambda_B)_{j} (\lambda_L)_{ \left[ (N-j+i)\, \!\!\!\!\!\!\!\ \mod\, \!\! N \right]} \label{conv}. 
}
For more details on the construction above see \cite{st}.

All the results we will give rely on the properties of the spectrum of Gram matrices of coherent states which we give collectively in the Appendix for convenience.
As this spectrum has roughly two regimes of behaviour, depending on the amplitudes $\alpha$ and $\beta$ being below or above one, we will separately address two distinct cases: small amplitude amplification (where $0<\alpha \leq \beta \leq 1$), and general amplification (all other amplitude combinations).
We begin by considering the scenario where both input and output amplitudes are small, i.e. less than one. 
From a practical standpoint, low amplitude amplification is of high importance since weak coherent states are often used in quantum information protocols. For sufficiently high amplitudes (also depending on $N$, that is, how many the  states are), the symmetric sets of coherent states  are effectively \textit{classical}, that is, mutually almost orthogonal, and can be reliably distinguished.
From a theoretical viewpoint, adhering to low amplitudes allows us to derive useful properties which do not hold for higher amplitudes.

\subsection{Small amplitude amplification}
\label{sec:smallamp}
If the amplitudes  $\alpha$ and  $ \beta$ of sets of symmetric coherent states $A$ and $B$, respectively, satisfy $\vert \alpha \vert < \vert \beta \vert < 1$, the following two properties hold for the spectra of their corresponding Gram matrices $G_A$ and $G_B$.

\noindent\textbf{Property 1}: the eigenvalues of 
$G_A$ appear in strictly decreasing order, where the order is induced by 
the order of the diagonal elements of the diagonalized matrix obtained by the conjugation of 
$G_A$ with the $uDFT$ matrix  (cf. Lemma \ref{lemord} below). 
 This does not hold for higher amplitudes.

\noindent\textbf{Property 2}: the quotient of the last eigenvalues of $G_A$ and $G_B$ is smaller than the quotient of any other two corresponding eigenvalues (cf. Lemma \ref{lemquot} below and the derivation preceding it). Again, this holds only in the small amplitude regime.

For proofs, please see the Appendix.

\textbf{Property 2} above implies that the upper bound on the optimal success probability $p_{up}$ in the low-amplitude regime, addressed in the beginning of this section, is reached in the leakless scenario, as we now show.
First, we note the link between the optimal success probability $d_S$ of uniformly and unambiguously discriminating a set of pure states $S$ and the spectrum of the Gram matrix $G_S$ of $S$: the optimal success probability $d_S$ is equal to the smallest eigenvalue of $G_S$ (this is easily derived from the results in \cite{chefles-2003, PhysRevA.65.052314} as was done in \cite{st}).
Also, the sufficient criterion (\ref{critEvs}) for the existence of a probabilistic leakless transform taking the states from $A$ to $B$ where both sets of states are symmetric, succeeding with the probability $p$, can be written as
\EQ{
\lambda_A - p \lambda_B \geq 0,
}
where $\lambda_A$ and $\lambda_B$ are the vectors of eigenvalues of matrices $G_A$ and $G_B$ as discussed in the previous section.
To see this, note that if the transform is leakless, then $\lambda_B \ast \lambda_L = \lambda_B$.
The maximal possible $p$ is then equal to $\min_j ({\lambda_A^j}/{\lambda_B^j})$, where $\lambda_A^j$ and $\lambda_B^j$ are the $j^{th}$ components of the vectors $\lambda_A$ and $\lambda_B$, respectively.
Now, by the second property, this minimum is attained for the last eigenvalues (i.e. $j=N-1$), which is exactly the upper bound $p_{up}$.
Thus, there exists a leakless transform saturating the upper bound of the success probability of amplification $p_{up}$.

Moreover, it can be shown by using \textbf{Property 1} that this bound is saturated \textit{only} by a leakless transform in the small amplitude regime.
From criterion (\ref{critEvs}), if there exists an amplification transform with a non-trivial leak, succeeding with some probability $p$, then the relation
 \EQ{
 \lambda_A - p \lambda_B \ast \lambda_{L} \geq 0
 }
holds, where $\lambda_{L}$ is the vector of eigenvalues of the Gram matrix of the leak.
Note that here we are assuming that the Gram matrix of the leak diagonalizes in the unitary discrete Fourier transform basis, which is justified without the loss of generality due to Lemma 4 in~\cite{st}.
If the leak is not trivial (not a fixed state) then $\lambda_{L}$ is a vector of non-negative numbers adding up to $N$, at least two of which are not zero.
Then note that the vector $\lambda_C=\lambda_B \ast \lambda_{L}$ contains the (normalized) weighted sums of the components of $\lambda_B$, the weights being the elements of  $\lambda_{L}$ (see the definition of the discrete convolution of vectors in expression (\ref{conv})).
Since the smallest component $\lambda_B^{min}$ is the unique last component of $\lambda_B$ (for $|\beta| <1$ by \textbf{Property 1}), and at least two of the elements of $\lambda_{L}$ are non-zero, the last component of   $\lambda_C$ is strictly greater than $\lambda_B^{min}$.
But then it holds that
\EQ{
p \leq \frac{\lambda_A^{N-1}}{\lambda_C^{N-1}} <\frac{\lambda_A^{N-1}}{\lambda_B^{min}}=\frac{\lambda_A^{min}}{\lambda_B^{min}}.
}

Hence, the success probability of any leaky (non-leakless) amplification transform for low amplitudes is strictly less than optimal.

Thus we have shown that small amplitude amplification can be done optimally, i.e. saturating the obvious upper bound of the success probability $p_{up}$, and that this optimal transform is always leakless.
The amplification procedure properties change significantly when one is interested in amplification involving states with amplitudes above unity, as we will see next.

\subsection{General amplification}

For `any amplitude' amplification, i.e. when $\beta >1$, we no longer have the convenient properties given in the previous subsection. In particular, optimal transforms can be leaky, in which case the upper bound $p_{up}$ derived through the probabilities of unambiguous discrimination (see expression (\ref{up})) sometimes no longer can be reached.
More formally, we have the following lemma:

\LE \label{nolf}
Let $\lambda_B^{min}$ be the smallest eigenvalue of the Gram matrix of the target, amplitude amplified, symmetric set of coherent states.
Then if $\lambda_B^{min}$ is a unique smallest eigenvalue then an optimal amplification transform with a non-trivial leak does not saturate the upper bound $p_{up}$.
\EL
\noindent\textbf{Proof:\\}
Let $c_j$ denote the $j^{th}$ component of the vector $\lambda_C=\lambda_B \ast \lambda_{L}$, where $\lambda_B$ and $\lambda_{L}$ are vectors of eigenvalues of the Gram matrices of the target states and the leak states. Let $c^{min} = \min_{j} c_j$.
Then, if  $\lambda_B^{min}$ is unique, and since  $\lambda_B \ast \lambda_{L}$ contains the (normalized) weighted sums of the components of $\lambda_B$, the weights being the elements of  $\lambda_{L}$, and at least two weights are not zero,  it holds that $\lambda_B^{min} < c^{min}$.

Let $p$ be the success probability of  an  optimal amplitude amplification transform with the leak characterized by $\lambda_{L}$. Then it holds that
 \EQ{
 \lambda_A - p \lambda_C \geq 0.
 }
Also, due to optimality, for some component $j$ it holds that
\EQ{
\lambda_{A}^j - p c_j = 0.
}
Assume first that $\lambda_{A}^j = \lambda_{A}^{min}$.
Then \EQ{
p = \frac{\lambda_{A}^{min}}{c_j}, }
and because $\lambda_B^{min} < c_{min}$ it holds that
 \EQ{
p = \frac{\lambda_{A}^{min}}{c_j} <   \frac{\lambda_{A}^{min}}{\lambda_B^{min}} = p_{up}, } so the upper bound is not saturated.

Assume now that $\lambda_{A}^j \not= \lambda^{A}_{min}=\lambda_A^l$ for some position $l \not = j.$
Since 
 \EQ{
 \lambda_A - p \lambda_C \geq 0
 }
it holds that 
\EQ{
p \leq \min_i \frac{\lambda_A^i}{c_i},
}
so since
\EQ{p = \frac{\lambda_{A}^{j}}{c_j}}
it holds that 
\EQ{
p = \frac{\lambda_{A}^{j}}{c_j} \leq  \frac{\lambda_{A}^{l}}{c_l} =  \frac{\lambda_{A}^{min}}{c_l} \leq \frac{\lambda_{A}^{min}}{c_{min}}<\frac{\lambda_{A}^{min}}{\lambda_{B}^{min}}=p_{up}.
}
Therefore the upper bound is not attained, and the lemma holds. \qed

With Lemma~\ref{nolf} in place, we now show through an example that in the case of general amplification, the leakless case may not be optimal, and the upper bound $p_{up}$ can sometimes no longer be obtained.
Consider amplification of a symmetric set of 4 coherent states from amplitude $\alpha = 2$ to amplitude $\beta = 2.3$.
The eigenvalues of the corresponding Gram matrices are then given by
\EQ{
\lambda_A = \left[ 0.976392, 0.971942, 1.02428, 1.02739 \right]^T\\
\lambda_B = \left[1.00553, 0.991527, 0.99452, 1.00842 \right]^T
}
and the upper bound is given with $p_{up} = 0.980248 .$
Note that the smallest eigenvalue of the Gram matrix of the target states is unique, so Lemma \ref{nolf} can be applied, and the upper bound cannot be reached in the leaky setting.

What remains to be seen is what the success probability of a leakless transform is.
The leak of a leakless transform are kets with only global phases possibly differing.  Lemma 4 in \cite{st} can still be applied in this case, hence we may assume that this leak is symmetric. This implies that the argument of the global phase of the $k^{th}$ ket is ``symmetric'' as well and will be of the form $\theta_k =  \pi k j /2 $ for $j = 0, \ldots, 3.$
By the properties of the discrete Fourier transform of powers of roots of unity, the vector of eigenvalues of such a leak will be a vector with all components zero, except at the position  $((4 - j \mod\,4)+1$ where its entry is 4.

 A convolution of a vector comprising zeroes, except at one position where the entry is one (or a constant $c$), with any other vector induces a circular permutation of the other vector (multiplied by the constant $c$).
Hence, we can directly check the optimal leakless success probability of the leakless amplification procedure, by going through all the circular permutations of $\lambda_B$. We find that the optimal leakless transform succeeds with probability $p_{leakless} = 0.977298<p_{up}$.
So, the upper bound cannot be reached for the leakless scenario either, which means that, surprisingly, it cannot be reached at all.
We note that although the values used in this analysis are numerical, the discrepancies the conclusion relies on (i.e. the uniqueness of the smallest eigenvalue and comparison of magnitude of the quotients) are well within numerical precision, hence the conclusion is unlikely to be a numerical artifact. 

Now, is there a leaky transform that does not saturate the bound, but does better than the best leakless transform? Using the optimization technique developed in \cite{st} we find that the success probability of an optimal transform for this example is $p_{opt} = 0.978604$ which is slightly larger than the optimal leakless transform $p_{leakless} = 0.977298$, and, necessarily, strictly below the upper bound $p=0.980248$.

To summarize, we have proven the following:
\begin{itemize}
\item The success probability of amplifying a symmetric set $A$ of $N$ coherent states of amplitude $\alpha$ to the states in a symmetric set $B$ of coherent states of a larger amplitude $\beta$, for small amplitudes $|\alpha| < |\beta| < 1$, can reach the upper bound imposed by the ratio of success probabilities of optimal unambiguous discrimination of sets $A$ and $B$, respectively.
\item For small amplitudes $|\alpha| < |\beta| < 1$ the optimal transform is always leakless.
\item The optimal success probability of amplification of small amplitudes is explicitly given by
\EQ{
p_{opt} = \frac{ \sum_{r=0}^{\infty} \frac{\alpha^{2(N (r+1)-1  )}}{(N (r+1)-1)!}}{ \sum_{r=0}^{\infty} \frac{\beta^{2(N (r+1)-1  )}}{(N (r+1)-1)!}}
.}
(please see the Equation (\ref{evsmin}) in 
the Appendix, and the subsequent paragraph).
\item If $\vert \beta \vert >1$, the numerical testing we have performed indicates that the
upper bound imposed by the ratio of the success probabilities for unambiguous discrimination of the states in sets $A$ and $B$ cannot always be reached, and optimal transforms may be leaky.
\end{itemize}

\section{Conclusions}

In this work, we have shown that truly noiseless amplification of coherent states is possible if one only requires the amplification to work perfectly for a finite number of of states. 
Similarly, perfect cloning of any other linearly independent states is also possible, and amplification is clearly closely related to cloning.
Depending on whether the amplitude of the amplified ``target" states are below or above one, the optimal success probability may be simply obtained, or require optimization techniques like the ones we have developed in~\cite{st}. The average gain is in principle unlimited, since it is possible to base the amplification on unambiguous state discrimination. In case of success, this allows us to prepare an amplified state with arbitrary high amplitude. If we require a finite level of gain, the optimal success probability is higher than for unlimited gain.
We have also explained how to implement truly noiseless amplification based on unambiguous state discrimination using only linear optics.

If we visualize the $N$ coherent states to be amplified as the spines in an umbrella in an $N$-dimensional space, then noiseless amplification of these states, which decreases their pairwise overlaps, may be thought of as ``opening the umbrella". Sometimes the optimal amplification procedures may result in extra ``leak" and ``redundancy" states, apart from the desired amplified states. The leak and the redundancy may be correlated with and therefore carry information about the input state.
Since the optimal ``umbrella transform" for truly noiseless amplification is always leakless when the amplitude of the  amplified (target) states is below one, as we have shown, this regime may be convenient if cryptographic aspects come into consideration. 
For example, in a two-party protocol, where Alice sends some quantum states to Bob who is supposed to further transform them, Alice can monitor the success probability declared by Bob. If it is optimal, she knows that there can be no additional leak (assuming that Alice uses some other way of checking that when Bob does declare that the process has succeeded, he has indeed obtained the quantum state he is supposed to). 
A related situation arises in blind quantum computing, where Alice wants to run a quantum computation on Bob's quantum computer without Bob learning about her data or her algorithm~\cite{10.1109/FOCS.2009.36}. In the original scheme, Alice is required to prepare single-qubit states. If Alice only can prepare, say, weak coherent states, then one possibility may be for Alice to require Bob to turn these into single-qubit states in such a way that Alice can monitor any additional information Bob may gain. Such transforms from symmetric coherent states to symmetric qubit states were considered in~\cite{st}. If the amplitude of the target states is above one, then the optimal ``umbrella" amplification transform may be leaky.

A few years ago, quantum cloning attracted widespread attention, see e.g.~\cite{WoottersZurek, BuzekHillery, Peresclon}.
Amplification and cloning are closely connected, especially for coherent states, since for example the state $|\alpha\rangle\otimes|\alpha\rangle$ may be transformed into $|\sqrt 2\alpha\rangle$ using a beam splitter, and vice versa. More generally, if $g=\sqrt N$, then the state $|g\alpha\rangle$ is equivalent to $N$ copies of $|\alpha\rangle$, in the sense that $|\sqrt N\alpha\rangle$ can be transformed into $N$ copies of $|\alpha\rangle$ (and vice versa) by a linear optical network (a balanced multiport). It is well known that perfect universal quantum cloning, i.e. of arbitrary states, is not possible~\cite{WoottersZurek, Peresclon}, but cloning with imperfect fidelity is. 

The fidelity of deterministic cloning can be improved if prior knowledge about the input states is available. Optimal cloning fidelity in the presence of prior knowledge for the case of cloning of CV systems, and in particular coherent states, has recently been addressed. Improvements have been shown, for instance, in settings where the input coherent states are picked from finite symmetric Gaussian distributions \cite{Cochrane, Sabuncu}, have a fixed phase and a wide spread of possible mean photon numbers \cite{Sabuncu,Guo2,Cochrane}, or have a fixed mean photon number (but an arbitrary phase) \cite{Sacchi, Muller}.
In \cite{Muller} it was shown that for the case of the latter type of prior knowledge -- the so-called phase covariant cloning \cite{Sacchi} -- fidelity of the output clones can be further improved if the cloning process is allowed to be probabilistic and heralded. However, perfect fidelity is only reached in the limits of zero success probability, and/or zero amplitude.
 
On the other hand, probabilistic perfect cloning of linearly independent states is possible~\cite{DuanGuo}. This mirrors the fact that probabilistic perfect amplification of linearly independent states is possible with finite success probability, as we have discussed.

To elaborate on the connection to cloning, the existing schemes for ``noiseless" probabilistic amplification of coherent states are (almost) perfect cloners for coherent states, but do not clone superpositions of coherent states as well. For example, choosing $g=\sqrt 2$, if $|\alpha\rangle \rightarrow |\sqrt 2\alpha\rangle$ for any $\alpha$, then  the ``cat" state $N_\pm(|\alpha\rangle\pm|-\alpha\rangle)$ would change into $N'_\pm(|\sqrt 2\alpha\rangle\pm|-\sqrt 2\alpha\rangle)$, which may be transformed into $N'_\pm(|\alpha\rangle\otimes|\alpha\rangle\pm|-\alpha\rangle\otimes|-\alpha\rangle)$ using a balanced beam splitter. This state 
is not equal to $N_\pm(|\alpha\rangle\pm|-\alpha\rangle)\otimes N_\pm(|\alpha\rangle\pm|-\alpha\rangle)$, that is, to two copies of the original cat state. This is similar to the simple proof that universal cloning is impossible~\cite{WoottersZurek}.

Nevertheless, this feature is not a disadvantage when perfect amplification schemes are used e.g. to enhance entanglement. That the operation $|\alpha\rangle \rightarrow |g\alpha\rangle$ only has unit fidelity in the limit of vanishing success probability, on the other hand, is a disadvantage.
If we select a finite linearly independent set of states $|\alpha_i\rangle$ which the amplification should work perfectly for, then the fidelity of the probabilistic process $N\sum_i c_i|\alpha_i\rangle \rightarrow N'\sum_i c_i|g\alpha_i\rangle$
can be truly perfect, as we have pointed out. The price we have to pay is that the scheme {\it must} be dependent on phase and amplitude. Nevertheless, since such schemes can be realized with only linear optics, as discussed in Sec.~\ref{sec:linopt}, we expect them to be of great interest for quantum information applications.

\section{Acknowledgements}
VD is fully and EA partially supported by EPSRC grant EP/G009821/1.

\appendix
\section{Properties of the spectrum of the Gram matrix of symmetric sets of coherent states}
\label{props}

The vector of eigenvalues of the Gram matrix of a symmetric set of coherent states $\lambda_{G_A}$ can be obtained by the discrete Fourier transform of the first row of $G_A$ (for details, see \cite{st}). Hence, the $j^{th}$ eigenvalue can be given as
\EQ{
\lambda_j = \sum\limits_{l=0}^{N-1} \exp\left(-2 j l \pi i /N    \right) \langle \alpha \ket{\alpha  \exp\left(2 l \pi i /N    \right)}.
}
Using the expansion of the coherent states in the Fock number basis the expression above can be written as
\EQ{
\lambda_j = 
\sum\limits_{l=0}^{N-1} \exp\left(-\frac{2 j l \pi i} {N}    \right) \sum_{r=0}^{\infty} e^{-\alpha^2} \frac{\alpha^{2r}}{r!} \exp \left(\frac{2 l r \pi i} {N}\right)
}
This can further be rearranged as follows:
\EQ{
\lambda_j 
&=&  e^{-\alpha^2} \sum\limits_{l=0}^{N-1} \sum_{r=0}^{\infty}  \exp\left(-2 j l \pi i/N    \right)\frac{\alpha^{2r}}{r!} \exp \left(2 l r \pi i/N\right)\nonumber\\
&=&e^{-\alpha^2}  \sum_{r=0}^{\infty} \frac{\alpha^{2r}}{r!}  \sum\limits_{l=0}^{N-1}  \exp\left(-2 j l \pi i /N    \right)  \exp (2 l r \pi i /n) \nonumber\\
&=&e^{-\alpha^2}  \sum_{r=0}^{\infty} \frac{\alpha^{2r}}{r!}  \sum\limits_{l=0}^{N-1}  \exp (2 l (r-j) \pi i /N), \label{absconv}
}
where to get to the the step (\ref{absconv}) we used the fact that the infinite sum is absolutely convergent, thereby allowing the commuting of sums.

By the properties of sums of roots of unity, the expression  $\sum\limits_{l=0}^{N-1}  \exp (2 l (r-j) \pi i /n)$ is equal to $n$ if $r-j$ is divisible by $N$ and zero otherwise.
Hence we obtain
\EQ{\label{evs}
\lambda_j = e^{-\alpha^2} N  \sum_{r=0}^{\infty} \frac{\alpha^{2(N r +j )}}{(N r +j)!}.  
}

The elements in the sum above appear as the summands in the Taylor expansion of $e ^ {2 \alpha}$. For any $j$ this sum collects  every $N^{th}$ summand from the Taylor series expansion starting from the $j^{th}$ summand. 
We note that the eigenvalues above can be expressed in a closed form in terms of generalized hypergeometric functions.
Using the presented form of the eigenvalues $\lambda_j$ we can show that for amplitudes below unity, the order of eigenvalues is monotonously decreasing:
\LE\label{lemord}
Let $A$ be the symmetric set of $N$ coherent states as defined in expression (\ref{source}). Let $\lambda_A$ be the vector of eigenvalues of the Gram matrix $G_{A}$ generated by taking the discrete Fourier transform of the first row of $G_A$.
If $\lambda_j$ is the $j^{th}$ component of $\lambda_A$, then for the real amplitude $\alpha \leq 1$ the eigenvalues in $\Lambda$ are decreasingly ordered:
$$
\lambda_j \geq \lambda_{j+1}.
$$
\EL

\noindent\textbf{Proof:\\}
We will show that $\lambda_j - \lambda_{j+1} \geq 0.$ By using expression (\ref{evs}) derived above, we obtain
\EQ{
&&\lambda_j - \lambda_{j+1}\\  &=& 
e^{-\alpha^2} N  \sum_{r=0}^{\infty} \frac{\alpha^{2(N r +j )}}{(N r +j)!}  - e^{-\alpha^2} N  \sum_{r=0}^{\infty} \frac{\alpha^{2(N r +j+1 )}}{(N r +j+1)!}\nonumber\\
&=& e^{-\alpha^2} N \alpha ^{2 j} \sum_{r=0}^{\infty} \frac{\alpha^{2 N r}}{(N r +j)!} \left(1 - \alpha ^2 \frac{1}{N r +j +1} \right),\nonumber
}
where the last step is possible due to absolute convergence of the sums above.
Note that the expression above is positive if $ \left(1 - \alpha ^2 \frac{1}{N r +j +1} \right)$ is positive.
It holds that $
{N r +j +1} \geq 1$, so for $\alpha \leq 1$ the expression above is positive and we have our claim.
Note also that in the case where $\alpha$ is strictly less than unity and positive, $\lambda_j$ is strictly greater than $\lambda_{j+i}$.
So, for amplitudes below $1$, the probability of success of unambiguous discrimination of symmetric sets of coherent states is given by the last eigenvalue in the vector $\lambda_A$. This eigenvalue is given by
\EQ{
\lambda_{min}= e^{-\alpha^2} N  \sum_{r=0}^{\infty} \frac{\alpha^{2(N (r+1)-1  )}}{(N (r+1)-1)!}. \label{evsmin}
}
In the case \textbf{Property 2} holds, from the equation above we can give the explicit optimal success probability of amplification of a set of symmetric coherent states. This is simply the quotient of the respective values of $\lambda_{min}$ for the two amplitudes, in the low amplitude regime.

In the remainder of the Appendix we prove \textbf{Property 2} from the main body of text.
Let $\lambda_{j}(\alpha)$ be the $j^{th}$ eigenvalue of the Gram matrix of the symmetric set of $N$ coherent states of (real) amplitude $\alpha.$  
\textbf{Property 2} states that
\EQ{
 \frac{\lambda_{j}(\alpha)}{\lambda_{j}(\beta)} \geq   \frac{\lambda _{N-1}(\alpha)}{\lambda_{N-1}(\beta)}
}
for all $j = 0,\ldots, N-1, $ and $0< \alpha<\beta<1$.
Since all the eigenvalues are positive and non-zero, the inequality above can be rewritten as
\EQ{
\frac{\lambda_j(\alpha)}{\lambda_{N-1}(\alpha)} \geq   \frac{\lambda_j(\beta)}{\lambda_{N-1}(\beta)}
}
which holds iff 
 ${\lambda_{j}(x)}/{\lambda_{N-1}(x)}$ 
is a decreasing function on $
\left(0, 1 \right)$.
Note that the functions $\lambda_{j}(x)$ are non-negative for all $j$ on the interval of interest. If it is the case that
${\lambda_j(x)}/{\lambda_{j+1}(x)}$
is a decreasing function on the interval $\left( 0, 1 \right)$ for all $j= 0, \ldots , N-2$, then the function $ {\lambda_{j}(x)}/{\lambda_{N-1}(x)}$ is decreasing as well, which would imply \textbf{Property 2}.
To see this, note that the equality 
\EQ{
 \frac{\lambda_{j}(x)}{\lambda_{j+1}(x)}  \frac{\lambda_{j+1}(x)}{\lambda_{j+2}(x)} \cdots  \frac{\lambda_{N-2}(x)}{\lambda_{N-1}(x)} =   \frac{\lambda_{j}(x)}{\lambda_{N-1}(x)}
}
holds for every $j$, and since the left-hand side of the expression above is a product of positive decreasing functions, the right-hand side must also be a decreasing function.
Hence, it will suffice to show that $\lambda_{j}(x)/{\lambda_{j+1}(x)} $ is a decreasing function on the interval of interest, which we state as the following Lemma.
\LE\label{lemquot}
The quotient of eigenvalues 
\EQ{
 \frac{\lambda_{j}(x)}{\lambda_{j+1}(x)} 
}
is a decreasing function on  $\left( 0, 1 \right)$ for all $j = 0, \ldots, N-2$.
\EL
\noindent\textbf{Proof:\\}
By recalling the analytic expression for the eigenvalues, given in (\ref{evs}), we have
\EQ{
  \frac{\lambda_{j}(x)}{\lambda_{j+1}(x)} &=& \frac{e^{-x^2} N  \sum_{r=0}^{\infty} \dfrac{x^{2(N r +j )}}{(N r +j)!}}{e^{-x^2} N  \sum_{r=0}^{\infty} \dfrac{x^{2(N r +j+1 )}}{(N r +j+1)!}}\nonumber \\ 
  &=&\frac{  \sum_{r=0}^{\infty} \dfrac{x^{2(N r +j )}}{(N r +j)!}}{  \sum_{r=0}^{\infty} \dfrac{x^{2(N r +j+1 )}}{(N r +j+1)!}}.
  }
Let us introduce the notation
$$
l_{j}(x) = \sum_{r=0}^{\infty} \frac{x^{2(N r +j )}}{(N r +j)!}.
$$
To prove lemma \ref{lemquot} we then need to show that 
$
{l_{j}(x)}/{l_{j+1}(x)}
$
is a decreasing function on  $\left( 0, 1 \right)$ for all $j = 0, \ldots, N-2$.
Note that the functions $l_{j}(x)$ are positive, strictly increasing and infinitely differentiable functions. Also, using the same technique we applied to prove the analogous property for the eigenvalues themselves, it holds that $l_{j}(x) \geq l_{j+1}(x) $ for all $j=0, \ldots N-2,$ and for $x \in \left( 0, 1 \right).$
Then, the quotient 
$
{l_{j}(x)}/{l_{j+1}(x)}
$
is decreasing in $x$ if and only if the derivative of the quotient over $x$ is non-positive on the interval of interest:
$$
\frac{l_{j}^\prime (x)l_{j+1}(x) -   l_{j}(x)l_{j+1}^\prime (x)       }{(l_{j+1}(x))^2} \leq 0
$$
Since the denominator of the fraction above is always positive, this inequality holds if and only if the inequality
$$
l_{j}^\prime (x)l_{j+1}(x) -   l_{j}(x)l_{j+1}^\prime (x)        \leq 0
$$
holds.

It is easy to verify the following property of the derivatives of the functions $l_j(x)$: 
\EQ{
l_j^\prime(x) =\frac{d}{dx} l_j(x) =2 x l_{j-1\, mod\, N}(x).
}
Hence we have
\EQ{
&&l_{j}^\prime (x)l_{j+1}(x) -   l_{j}(x)l_{j+1}^\prime (x)  \nonumber \\ 
&&~~= 2 x \left( l_{j-1\, mod\, N} (x)l_{j+1}(x) -   l_{j}(x)l_{j} (x) \right)
}
which is non-positive on the interval of interest if and only if 
\EQ{
 l_{j-1\, mod\, N} (x)l_{j+1}(x) -   l_{j}(x)l_{j} (x) \leq 0. \label{condition2}
}
Note that if $j=0$ the expression above resolves to
\EQ{
 l_{N-1} (x)l_{1}(x) -   l_{0}(x)l_{0} (x) \leq 0. 
}
Since $l_{N-1} (x) \leq l_{0}(x) $ and  $l_{1} (x) \leq l_{0}(x) $ on the interval $\left( 0, 1 \right)$, and since all the values these functions attain are positive, 
we have that for $j=0$ the condition given in expression (\ref{condition2}) holds.
By using the definitions of the functions $l_j(x)$, for   $j = 1, \ldots N-2$, we obtain
\begin{widetext}
\EQ{
 &&l_{j-1} (x)l_{j+1}(x) -   l_{j}(x)l_{j} (x) =   \sum_{r=0}^{\infty} \frac{x^{2(N r + j-1 )}}{(N r +j-1)!}  \sum_{r=0}^{\infty} \frac{x^{2(N r +j+1 )}}{(N r +j+1)!} - \sum_{r=0}^{\infty} \frac{x^{2(N r +j )}}{(N r +j)!}  \sum_{r=0}^{\infty} \frac{x^{2(N r +j )}}{(N r +j)!} \\
&&= x^{4 j} \sum_{r=0}^{\infty} (x^{2N})^r \frac{1}{(N r +j-1)!}  \sum_{r=0}^{\infty} (x^{2N})^r \frac{{1}}{(N r +j+1)!} - x^{4 j} \sum_{r=0}^{\infty}  (x^{2N})^r \frac{1}{(N r +j)!}  \sum_{r=0}^{\infty}  (x^{2N})^r \frac{1}{(N r +j)!}.
 }
The sign of the expression above is then equal to the sign of the expression
\EQ{
\sum_{r=0}^{\infty} (x^{2N})^r \frac{1}{(N r +j-1)!}  \sum_{r=0}^{\infty} (x^{2N})^r \frac{{1}}{(N r +j+1)!} -  \sum_{r=0}^{\infty}  (x^{2N})^r \frac{1}{(N r +j)!}  \sum_{r=0}^{\infty}  (x^{2N})^r \frac{1}{(N r +j)!} \label{finalexpr}.
 }
Note that to prove that $l_{j-1} (x)l_{j+1}(x) -   l_{j}(x)l_{j} (x) \leq 0$ for $j>0$ (and consequently \textbf{Property 2}),
it will suffice that the expression (\ref{finalexpr}) is negative  for all $x \in \left( 0, 1 \right),$ and for $j = 1 ,\ldots N-2$.
Also, since any positive power is a bijection on the interval $x \in \left( 0, 1 \right)$, and we require negativity on the entire interval, the expression(\ref{finalexpr}) is negative if and only if the expression  
\EQ{
\sum_{r=0}^{\infty} (x^{N})^r \frac{1}{(N r +j-1)!}  \sum_{r=0}^{\infty} (x^{N})^r \frac{{1}}{(N r +j+1)!} -  \sum_{r=0}^{\infty}  (x^{N})^r \frac{1}{(N r +j)!}  \sum_{r=0}^{\infty}  (x^{N})^r \frac{1}{(N r +j)!}\label{criterionF}
 }
is negative on the same interval.
 \end{widetext}
 
Consider now the family of functions
\EQ{
f_j(x) = \sum_{r=0}^{\infty} \frac{x^{(N r +j )}}{(N r +j)!}. \nonumber
}
Using the same construction as for the functions $l_j(x)$, it is easy to see that 
${f_j(x)}/{f_{j+1}(x)}$
is a decreasing function on $\left( 0, 1 \right)$ for $j = 1, \ldots N-2$ if and only if the expression (\ref{criterionF}) is negative on the same interval.
For these functions $f_j$ it is also easy to see that they are positive, strictly increasing, infinitely differentiable, 
and $f_j(x)\geq f_{j+1}(x)$ holds on the interval of interest for $j=0, \ldots N-2.$
It also holds that 
 \EQ{
 \frac{d}{dx} f_j(x) = f_{j-1\, mod\, N}(x).
 }

Recall the property of log-concavity: a function is log-concave (on an interval) if the logarithm of that function is concave on the same interval.
For functions which are twice differentiable, log-concavity holds if and only if the quotient of the derivative of the function and the function itself is decreasing (on the same interval).
Hence, the requirement that 
${f_j(x)}/{f_{j+1}(x)}$
is decreasing on the interval of interest is equivalent to the requirement that $f_{j+1}(x)$ is a log-concave function.

Here we invoke the following result given in Lemma 1 of the manuscript \cite{LCLC3}, also a consequence of the Lemma 3 in the Appendix of \cite{LCLC} (a published version of the aforementioned manuscript):
\LE
Let $g(x)$ be a strictly monotonic, twice differentiable function on the interval $( a, b )$. Let also $g(a)=0$ or $g(b)=0$.
Then if the derivative $g^\prime (x)$ is log-concave on the same interval,  $g(x)$ is log-concave on the interval.
\EL

Since for all $j>0$ the function $f_{j}(0)$ is zero, and all the functions $f_j$ are strictly increasing, it holds that $f_{j+1}(x)$ is log-concave if $f_{j}(x)$ is log-concave.
Inductively, if $f_1(x)$ is log-concave, so is $f_j(x)$ for all $j=2, \ldots, N-1.$ 
To finish the proof of Lemma \ref{lemquot} and thus of \textbf{Property 2}, we finally need to show that $f_1(x)$ is log-concave on $ \left( 0, 1 \right)$.
Recall that $f_1(x)$ is log-concave on the interval of interest if the quotient
${f_0(x)}/{f_{1}(x)}$
is decreasing on the interval.
This holds if the inequality
\EQ{
&&f^\prime_0(x) f_{1}(x) - f^\prime_1(x)f_0(x) \nonumber \\ 
&&~~=  f_{N-1}(x) f_{1}(x) - f_0(x)f_0(x) \leq 0 \nonumber
}
holds. But since we have that $f_j(x)\geq f_{j+1}(x)$ holds on the interval of interest for $j=0, \ldots N-2$ and since all the functions above attain positive values, this inequality is satisfied. Hence  Lemma \ref{lemquot} and \textbf{Property 2} are proven. 

We note that the functions $l_j(x)$ and $f_j(x)$ are sub-series of the Taylor expansion of the functions $e^{x^2}$ and $e^{x}$ about the point $x=0$, respectively, and as such are absolutely convergent, which allows the unrestricted reshuffling of sums.

\end{document}